\documentclass{book}
\usepackage{makeidx,epsfig}
\usepackage{amsthm,amsmath,amssymb}
\usepackage{setspace,graphicx}
\usepackage{Generic,comment}
\usepackage[sort,longnamesfirst]{natbib}

\usepackage{url,doi}

\usepackage{algorithm}
\usepackage{algpseudocode}

\newcommand{\half}{\mbox{\small $\frac{1}{2}$}}

\numberwithin{equation}{section}
\theoremstyle{plain}

\usepackage[normalem]{ulem}
\usepackage[dvipsnames]{xcolor}

\newcommand{\reallyzap}[1]{}

\begin{document}

\setcounter{chapter}{15}
\counterwithout{section}{chapter}
\graphicspath{{./}{Figs/}}

\chapter[Computationally intensive inverse problems]{Posterior exploration for computationally
intensive forward models}

\begin{center}
\begin{large}
{\bf Mikkel B. Lykkegaard, Colin Fox, Dave Higdon, C. Shane Reese,\\ and J. David Moulton}
\end{large}\\[2em]
\end{center}
{\em To appear in the Handbook of Markov Chain Monte Carlo (2\textsuperscript{nd} edition),\\ probably as Chapter 16.}


\section{Introduction}
In a common inverse problem, we wish to infer about an unknown spatial
field $x=(x_1,\ldots,x_m)^T$ given indirect observations $y =
(y_1,\ldots,y_n)^T$.  The observations, or data, are linked to the
unknown field $x$ through some physical system
\[
 y = \zeta(x) + \epsilon
\]
where $\zeta(x)$ denotes the actual physical system and $\epsilon$ is an
$n$-vector of observation errors.  Examples of such problems include
medical imaging \citep{kaip:some:2004}, geologic and hydrologic
inversion \citep{stenerud:aam}, and cosmology \citep{jimenez2004fcp}.
When a forward model, or simulator $\eta(x)$, is available to model the physical process, one can model the data using the simulator
\[
  y = \eta(x) + e,
\]
where $e$ includes observation error as well as error due to the fact
that the simulator $\eta(x)$ may be systematically different from
reality $\zeta(x)$ for input condition $x$.  Our goal is to use the
observed data $y$ to make inference about the spatial input parameters
$x$ -- predict $x$ and characterize the uncertainty in the prediction
for $x$.

The likelihood $L(y|x)$ is then specified to account for both mismatch
and sampling error. We will assume zero-mean Gaussian errors so that
\begin{equation}
\label{eq:like}
 L(y|x) \propto \exp\{-\half (y - \eta(x))^T \Sigma_e^{-1} (y-\eta(x)) \},
\end{equation}
with $\Sigma_e$ known.  It is worth noting that the data often
come from only a single experiment. So while it is possible to
quantify numerical errors, such as those due to discretization
(e.g.,\cite{kaip:some:2004,nissinen2008mst}),
there is no opportunity to
obtain data from additional experiments for which some controllable
inputs have been varied.  Because of this limitation, there is little 
hope of determining the sources of error in $e$ due to model
inadequacy.  Therefore, the likelihood specification will
often need to be done with some care, incorporating the modeler's
judgment about the appropriate size and nature of the mismatch term.

In many inverse problems we wish to reconstruct $x$, an unknown
process over a regular 2-d lattice.  We consider systems for which the
model input parameters $x$ denote a spatial field or image.
The spatial prior is specified for $x$, $\pi(x)$ which typically takes
into account modeling, and possibly computational considerations.

The resulting posterior is then given by 
\[
\pi(x|y) \propto L(y|\eta(x)) \times \pi(x).
\]
This posterior can, in principle, be explored via Markov chain Monte
Carlo (MCMC). However the combined effects of the high dimensionality
of $x$ and the computational demands of the simulator make
implementation difficult, and often impossible, in practice. By
itself, the high dimensionality of $x$ isn't necessarily a
problem. MCMC has been carried out with relative ease in large image
applications \citep{weir:1997,rue2001fsg}.  However, in these
examples, the forward model was either trivial, or
non-existent. Unfortunately, even a mildly demanding forward
simulation model can greatly affect the feasibility of doing MCMC to
solve the inverse problem.

In this chapter we apply a standard single-site updating scheme that
dates back to \citet{Metrop:53} to sample from this posterior.  While
this approach has proven effective in a variety of applications, it
has the drawback of requiring millions of calls to the
simulation model.  In Section \ref{sec:mvmcmc} we consider an MCMC
scheme that uses highly multivariate updates to sample from
$\pi(x|y)$: the multivariate random walk Metropolis algorithm
\citep{gelman1996emj}.  
%
Such multivariate updating schemes
are alluring for computationally demanding inverse problems since they
have the potential to update many (or all) components of $x$ at once,
while requiring only a single evaluation of the simulator.  

Next, in
Section \ref{sec:approx}, we consider augmenting the basic posterior
formulation with additional formulations based on faster, approximate
simulators.  
We create two faster, approximate forward models: $\eta_\text{m}(x)$ by leveraging an incomplete multigrid solve; and $\eta_\text{c}(x)$ by coarsening
the initial ``fine-scale'' formulation described in the next section.  
%
These approximate simulators can be used in a delayed acceptance scheme
\citep{liu2001mcs,christen2005mcm}, as well as in an augmented
formulation \citep{higd:lee:bi:2002}.  Both of these recipes can be
utilized with any of the above MCMC schemes, often leading to
substantial improvements in efficiency.  
For a given $x$, resulting output produced by the approximate multigrid model $\eta_\text{m}(x)$ is quite close to that obtained by exact solve of the fine scale model $\eta(x)$.  However, the coarser model $\eta_\text{c}(x)$ shows substantial, systematic differences from the output of the fine-scale model $\eta(x)$.  So we also consider the stochastic approaches for correcting systematic errors between the coarse- and fine-scale model output, introduced in~\citet{cui2011bayesian}. We find that the corrected coarse model enables cheap and semi-automatic generation of multivariate proposals from the single-site proposal, with an appreciable acceptance rate in the fine-scale chain, giving improved efficiency.

The updating schemes are illustrated with an electrical impedance tomography (EIT) application described in the next section, where the values of $x$
denote electrical conductivity of a 2-d object.  The chapter concludes
with a discussion and some general recommendations.

\section{An inverse problem in electrical impedance tomography}
\label{sec:eit}

\begin{figure}[h!t]
  \begin{center}
    \parbox[c]{0.36\linewidth}{%
    \includegraphics[width=0.84\linewidth]{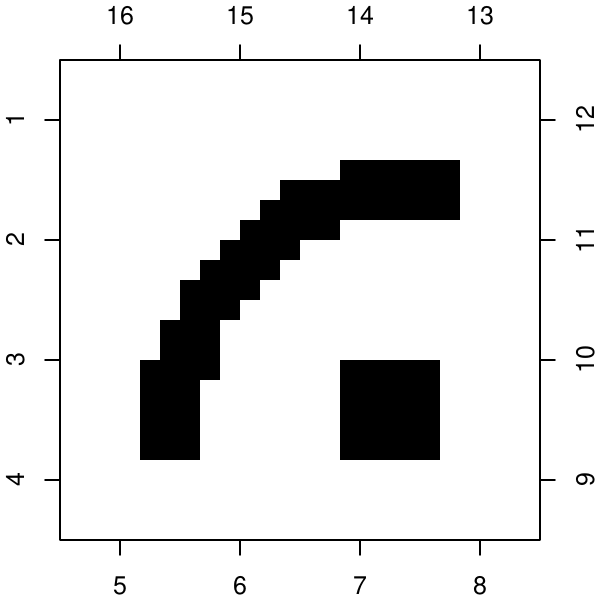}
    }
    \parbox[c]{0.63\linewidth}{%
    \includegraphics[width=1.0\linewidth]{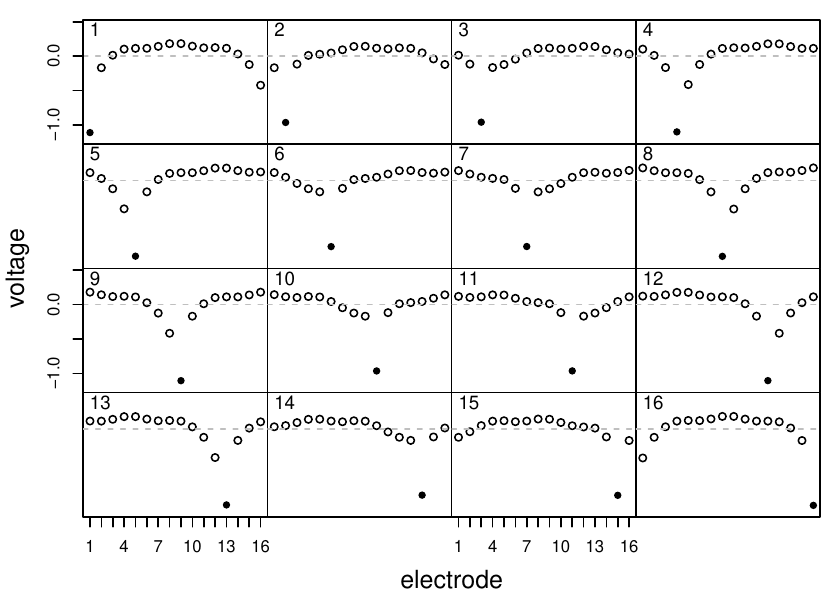}
    }
  \end{center}
  \caption{\label{fig:eit} A synthetic EIT application.  A 2-d object
    is surrounded by electrodes at 16 evenly spaced locations around
    its edge.  The conductivity of the object is 3 in the white
    regions, and 4 in the black regions (the units are arbitrary since
    the data are invariant to scaling of the conductivity).  First, a
    current of $I$ is injected at electrode 1, and extracted evenly at
    the other 15 electrodes.  The voltage is measured at each
    electrode, with respect to the mean voltage on electrodes.  
    This data is shown in the plot labeled 1 on the right.
    Similar experiments are carried out with each electrode taking a
    turn as the injector.  The resulting voltages are shown in the
    remaining 15 plots.  In each plot, the voltage corresponding to
    the injector electrode is given by a black plotting symbol.}
\end{figure}

Bayesian methods for EIT applications have been described in
\citet{fox1997sci}, \citet{kaipio2000sia} and \citet{andersen2003big}.
A notional inverse problem is depicted in Figure \ref{fig:eit}; this
setup was given previously in \citet{moul:fox:svya:2007}.  Here a 2-d
object composed of regions with differing electrical conductivity is
interrogated by $16$ electrodes.  From each electrode, in turn, a
current $I$ is injected into the object and taken out at a rate of
$I/(16-1)$ at the remaining 15 electrodes.  The voltage is then
measured at each of the 16 electrodes.  These 16 experimental
configurations result in $n=16 \times 16$
voltage observations which are denoted by the $n$-vector $y$. The
measurement error is simulated by adding iid mean 0 Gaussian noise to
each of the voltage measurements.  The standard deviation $\sigma$ of
this noise is chosen so that the signal to noise ratio is about
1000:3, which is typical of actual EIT measurements.  The resulting
simulated data is shown in the right frame of Figure \ref{fig:eit} --
one plot for each of the 16 circuit configurations.  In each of those
plots, the injector electrode is denoted by the black plotting symbol.

We take $s$ to denote spatial locations within the object $\Omega =
[0,1] \times [0,1]$, and take $x(s)$ to denote the electrical
conductivity at site $s$.  We also take $v(s)$ to be the potential at
location $s$, and $j(s)$ to be the current at boundary location $s$.
A mathematical model for the measurements is then the Neumann
boundary-value problem
\[
\begin{array}
[c]{rlll}%
{-\nabla}\cdot x\left(  s\right)  \nabla v \left(  s \right) &
= & 0 &
\qquad s\in\Omega\\
\displaystyle x\left(  s\right)  \frac{\partial v \left(
s \right) }{\partial n \left(  s\right)  } & = & j\left(  s\right) &
\qquad s\in
\partial\Omega ,
\end{array}
\]
where $\partial \Omega$ denotes the boundary of the object 
$\Omega$ and $n(s)$ is the unit normal vector at the boundary
location $s \in \partial \Omega$.
The conservation of current requires that the sum of the currents at
each of the 16 electrodes be 0. 
The arbitrary additive offset in the solution of this Neumann problem is determined by requiring that the sum of voltages at boundary electrodes is zero.

In order to numerically solve this problem for a given set of currents
at the electrodes and a given conductivity field, $x(s)$, the
conductivity field is discretized into an $m = 24 \times 24$ lattice.
We use a standard Cholesky-based sparse solver for this system.  Later in Section \ref{sec:approx}, we also create a faster, approximate solver based on an $8 \times 8$ discretization of the conductivity field.  

Now, for any specified conductivity configuration $x$ and current
configuration, the solver produces 16 voltages.  For all 16
current configurations, the forward solver produces an $n = 256$-vector
of resulting voltages $\eta(x)$. Hence the sampling model for the data
$y$ given the conductivity field $x$ is given by (\ref{eq:like}) where
$\Sigma_e = \sigma^2 I_n$.

\begin{figure}[h!t]
  \centerline{
   \input{mrf2d.pic}
   \includegraphics[width=2.0in,angle=0] {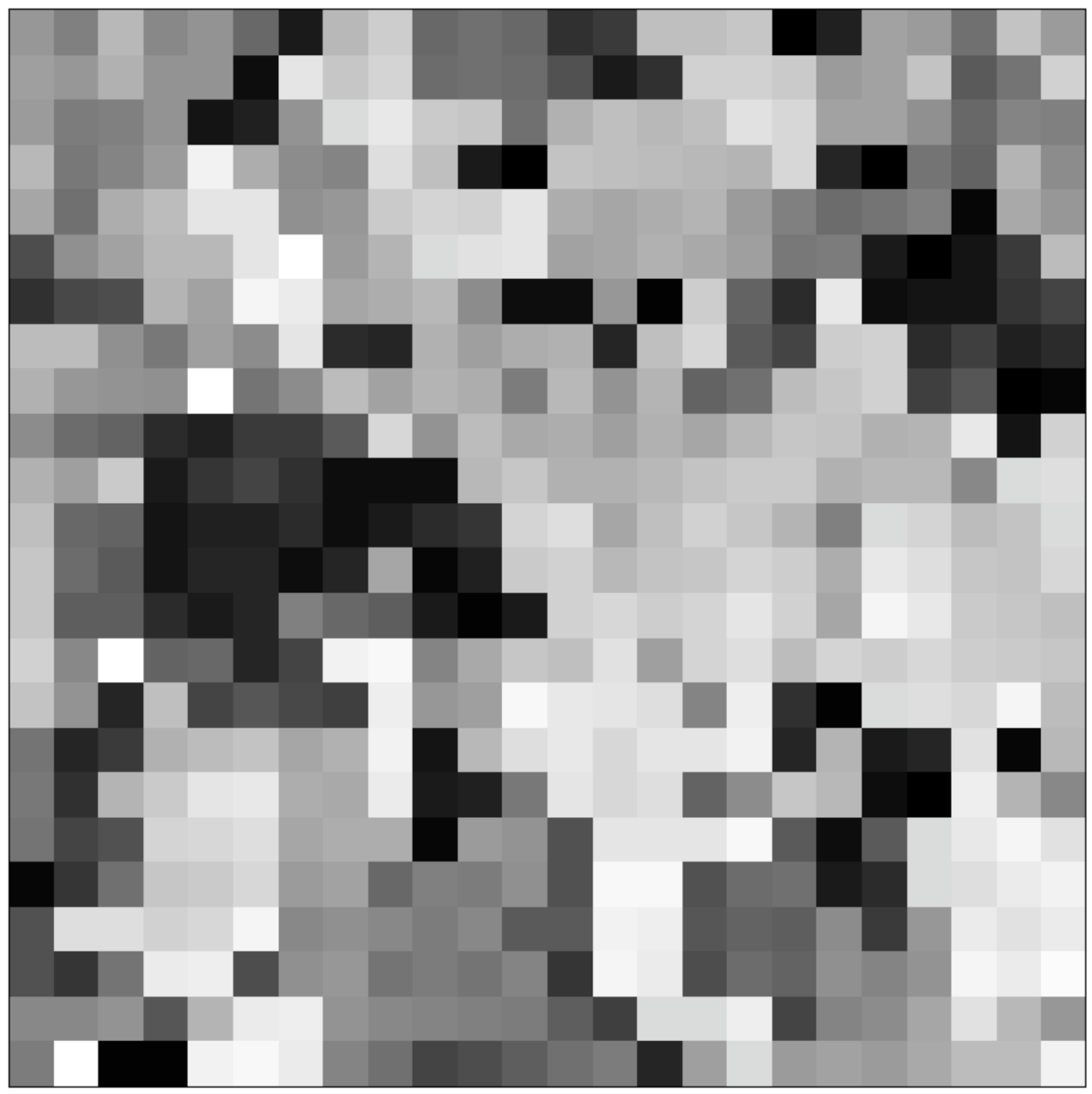}
  }
  \caption{\label{fig:prior} Left: the first order neighborhood MRF
    graph corresponding to the prior in (\ref{eq:grayprior}); each
    term in the sum corresponds to an edge in the MRF graph. Right: a
    realization from this gray level prior.}
\end{figure}

For the conductivity image prior, we adapt a Markov random field (MRF) prior
from \citet{gema:mccl:1987}.  This prior has the form
\begin{equation}
\label{eq:grayprior}
 \pi(x) \propto \exp\left\{ \beta \sum_{i \sim j} u(x_i - x_j) \right\},
        \, x \in [2.5,4.5]^m
\end{equation}
where $\beta$ and $s$ control the regularity of the field, and
$u(\cdot)$ is the tricube function of \citet{clev:1979}
\[
 u(d) = \left\{
      \begin{array}{cl} \displaystyle\frac{1}{s}(1 - \left|d/s \right|^3)^3 & \mbox{ if } -s < d <s\\
                              0 & \mbox{ if } |d| \geq s \, .
             \end{array} \right.
\]
The sum is over all horizontal and vertical nearest neighbors denoted
by $i \sim j$ and given by
the edges in the Markov random field graph in Figure \ref{fig:prior}.  Hence
this prior encourages neighboring $x_i$'s to have similar values, but
once $x_i$ and $x_j$ are more than $s$ apart, the penalty does not
grow.  This allows occasional large shifts between neighboring $x_i$'s.
For this chapter, we fix $(\beta,s) = (.5,.3)$.  A realization from
this prior is shown in the right frame of Figure \ref{fig:prior}.  A
typical prior realization shows patches of homogeneous values,
along with abrupt changes in intensity at each patch boundary.  This prior
also allows an occasional, isolated, extreme single pixel value. 

The resulting posterior density has the form
\begin{equation}
\label{eq:graypost}
\pi(x|y) \propto
\exp\left\{ -\frac{1}{2 \sigma^2} (y - \eta(x))^T  (y-\eta(x)) \right\}
\times
\exp\left\{ \beta \sum_{i \sim j} u(x_i - x_j) \right\},\, x \in [2.5,4.5]^m.
\end{equation}
The patchiness and speckle allowed by this prior along with the rather
global nature of the likelihood make posterior exploration for this
inverse problem rather challenging, and a good test case for various
MCMC schemes that have been developed over the years.  We note that nature of
the posterior can be dramatically altered by changing 
the prior specification for $x$.  This is discussed later in this section.

This chapter considers a number of MCMC approaches for sampling
from this posterior distribution.  We start at the beginning.

\subsection{Posterior exploration via single-site Metropolis updates}

A robust and straightforward method for computing samples from the
posterior $\pi(x|y)$ is the single-site Metropolis scheme, originally
carried out in \citet{Metrop:53} on the world's first computer with
addressable memory, the MANIAC.  A common formulation of this
scheme is summarized in Algorithm~\ref{alg:ssm} using pseudo code.
\begin{algorithm}[h!]
\caption{Single Site Metropolis}
\label{alg:ssm}
\begin{algorithmic}[1]
\Statex
\State initialize $x$
\For{$k=1:\text{niter}$}
  \For{$i=1:m$}
    \State $x' = x$
    \State $x_i' = x_i + z$, where $z \sim N(0,\sigma^2_z)$
    \If{ $u < \frac{\pi(x'|y)}{\pi(x|y)}$, where $u \sim U(0,1)$ }  \label{alg:ssm:ratio}
     \State set $x_i = x_i'$
   \EndIf
   \EndFor
\EndFor
\end{algorithmic}
\end{algorithm}\\[0cm]
This scheme is engineered to maintain  balance with the posterior distribution $\pi(x|y)$  -- so that the nett  movement between any two states $x$ and $x^*$ is done in
proportion to the posterior density at these two points. The width of
the proposal distribution $\sigma_z$ should be adjusted so that
inequality in line \ref{alg:ssm:ratio} is satisfied roughly half the time
\citep{gelman1996emj}, but an acceptance rate between 70\% and 30\%
does nearly as well for single-site updates.  
The single-site Metropolis in Algorithm~\ref{alg:ssm} scans through each of the parameter elements
in a fixed, deterministic order (for loop, steps 3--9) to define one \emph{sweep} over parameter values. 
One typically records the current value of $x$ after each sweep; we do so every 10 sweeps.

This single-site scheme was originally intended for distributions with
very local dependencies within the elements of $x$ so that the ratio
in line 6 simplifies dramatically.  In general, this simplification
depends on the full conditional density of $x_i$
\[
  \pi(x_i|x_{-i},y),\mbox{ where } x_{-i} = (x_1,\ldots,x_{i-1},x_{i+1},\ldots,x_n)^T.
\]
This density is determined by keeping all of the product terms in
$\pi(x|y)$ that contain $x_i$, and ignoring the terms that don't.
Hence the ratio in line \ref{alg:ssm:ratio} 
can be rewritten as
\[
   \frac{\pi(x'|y)}{\pi(x|y)} = \frac{\pi(x_i'|x_{-i},y)}{\pi(x_i|x_{-i},y)}.
\]
In many cases this ratio becomes trivial to compute.  However, in the
case of this particular inverse problem, we must still evaluate the
simulator to compute this ratio.  This is exactly what makes the MCMC
computation costly for this problem.

\begin{figure}[bh]
  \centerline{
   \includegraphics[width=5.5in,angle=0] {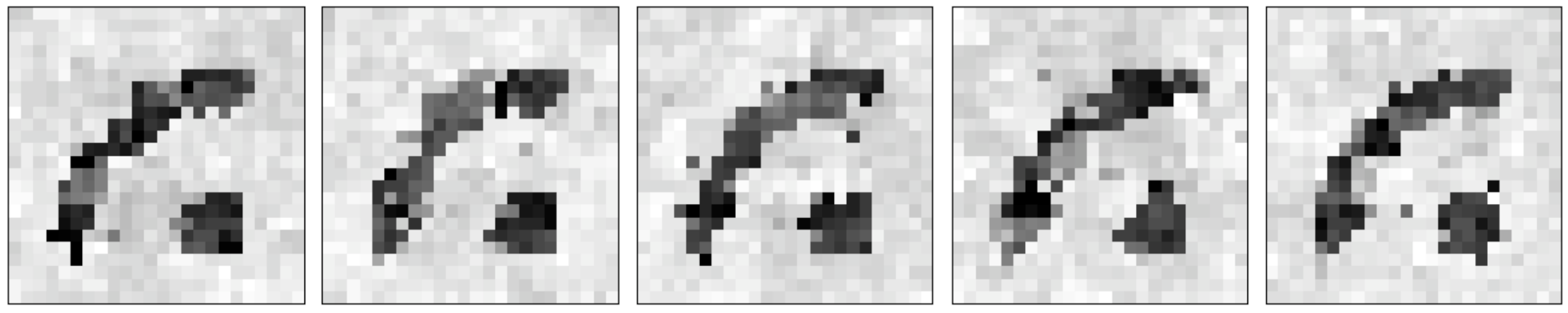}
  }
   \caption{\label{fig:realizations} Five realizations from the single-site Metropolis scheme.
   Realizations are separated by $1000$ sweeps over the $m$-dimensional image parameter $x$.}
\end{figure}
Nonetheless, this straightforward sampling approach does adequately
sample the posterior, given sufficient computational effort; see~\cite{FrigessiSingleSite} for a proof of ergodicity.  Figure \ref{fig:realizations}
shows realizations produced by the single-site Metropolis algorithm, separated
by $1000$ sweeps.
Inspection of these realizations makes it clear that 
posterior realizations yield a crisp distinction between the high and
low conductivity regions, as was intended by the MRF prior for $x$.
Around the boundary of the high conductivity region there is a fair amount
of uncertainty as to whether or not a given pixel has high or low
conductivity.

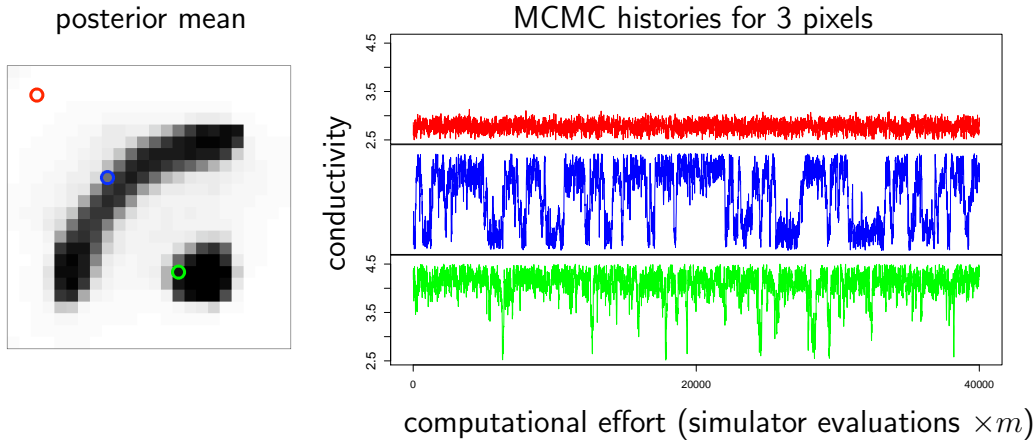
\begin{figure}[h!t]
\centerline{
\begin{picture}(385,170)(0,0)
\put(0,40){\includegraphics[width=1.5in,angle=0]{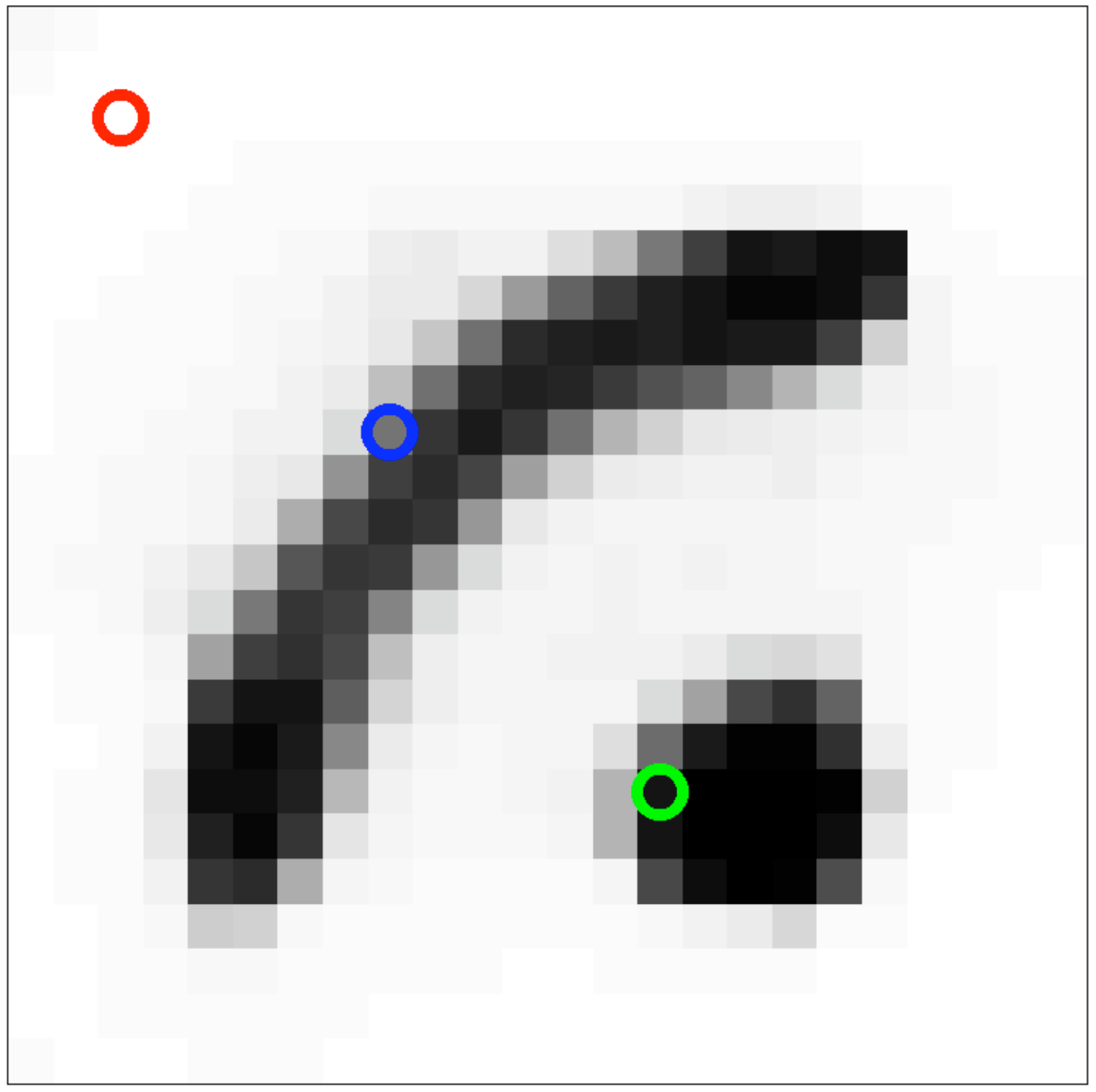}}
\put(130,20){\includegraphics[width=3.5in,angle=0]{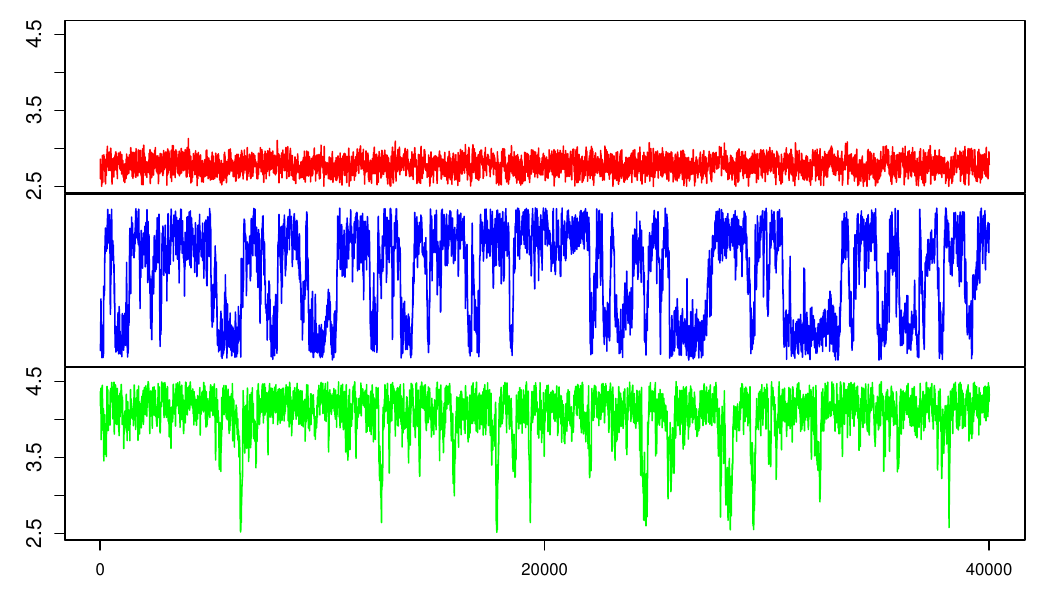}}
\put(270,12){\makebox(0,0){\sf computational effort (simulator evaluations $\times m$)}}
\put(120,70){\rotatebox{90}{\sf conductivity}}
\put(55,165){\makebox(0,0){\sf posterior mean}}
\put(260,165){\makebox(0,0){\sf MCMC histories for 3 pixels}}
\end{picture} 
}
\caption{\label{fig:ssm} Posterior mean image for $x$ and MCMC traces
  of three pixels: one which is predominantly light (small
  conductivity); one which is predominantly dark (high conductivity)
  and one which is on the edge of the object.  This MCMC run
  carries out $40000 \times m$ forward simulator evaluations.  The
  value of $x_i$ is given every 10th iteration (i.e. every $10\times
  m$ single-site updates).}
\end{figure}

Figure \ref{fig:ssm} shows the resulting posterior mean for $x$ and
shows the history of three pixel values over the course of the
single-site updating scheme.  The sampler was run until $40,000 \times
m$ forward simulations were carried out.  An evenly spaced sample of
4,000 values for three of the $m$ pixels is shown in the right frame of
Figure \ref{fig:ssm}.  Note that for the middle pixel (blue circle),
the marginal posterior distribution is bimodal -- some realizations
have the conductivity value near 3, others near 4.  Being able to move
between these modes is crucial for a well mixing chain.  Getting this
pixel to move between modes is not simply a matter of getting that one
pixel to move by itself; the movement of that pixel is accomplished by
getting the entire image $x$ to move between local modes of the
posterior.

\begin{figure}[t!h]
  \centerline{
   \includegraphics[width=4.0in,angle=0] {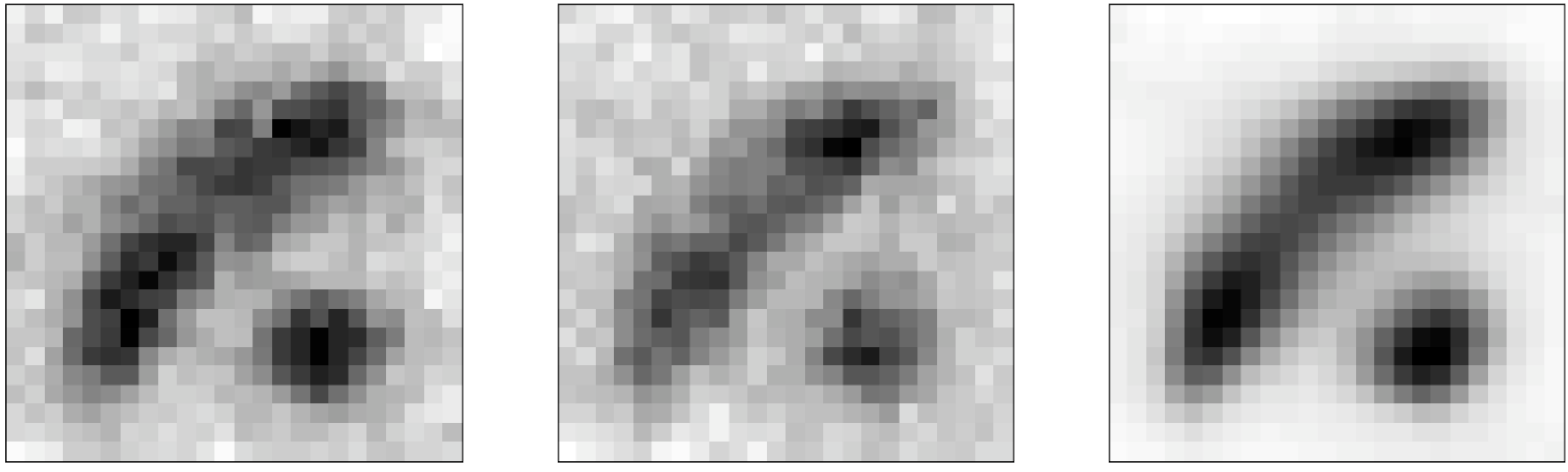}
  }
   \caption{\label{fig:gmrf} Two realizations and the posterior mean from the single-site Metropolis scheme
   run on the posterior resulting from the GMRF prior.
   Realizations are separated by $1000$ sweeps over the $m$-dimensional image $x$.}
\end{figure}

This local multimodality is largely induced by our choice of prior.
For example, if we alter the prior model in (\ref{eq:grayprior}) so that 
\begin{equation}
\label{eq:gmrf}
u(d) = -d^2, 
\end{equation}
we have a standard Gaussian Markov random field (GMRF) prior
for $x$.  If, in addition, the simulator is a linear mapping from inputs
$x$ to ouputs $\eta(x)$, the resulting 
posterior is necessarily Gaussian, and hence, unimodal.   While this is not 
true for nonlinear forward models/simulators, the GMRF prior still has substantial influence on
the nature of the posterior; indeed, recent theoretical results show that the posterior for EIT with GMRF prior is typically log-concave~\citep{Nikl-lecturenotes}, hence unimodal, confirming these computational observations.   Figure \ref{fig:gmrf} shows two realizations and the
posterior mean resulting from such a prior with $\beta=2$.  Here posterior realizations
are locally more variable -- the difference between neighboring pixels are generally
larger.  However the global nature of the posterior realizations are far more
controlled than those in Figure \ref{fig:realizations} since the GMRF prior
suppresses local modes that appear under the previous formulation.  This
resulting formulation is also far easier to sample, requiring about a 10th of
the effort needed for formulation in (\ref{eq:graypost}).  An alternate, controlling
prior formulation uses a process convolution prior for $x$ is given in the Appendix.
In addition to yielding a more easily sampled posterior, the prior also represents
the image $x$ with far fewer parameters than the $m$ used in the MRF specifications.


While these alternative specifications lead to simpler posterior distributions, they do so at the expense of
 overly smooth posterior realizations.  Still, such realizations may
be useful for exploratory purposes, and for initializing other samplers; we do not 
further pursue such formulations here. Instead, we focus on comparison of various MCMC schemes 
to sample the original gray level posterior in (\ref{eq:graypost}).  We use the
sample traces from the three pixels circled in Figure \ref{fig:ssm} to make comparisons
between a variety of samplers which are discussed in the next sections -- the movement of
these three pixels is representative of all the image pixels.  In particular,
we focus on the frequency of movement between high and low conductivity at these
sites.

\section{Multivariate updating schemes} \label{sec:mvmcmc}

Schemes that propose to update more than just a single component of
$x$ at a time have the potential to reduce the computational burden of
producing an MCMC sample from $\pi(x|y)$.  The single-site scheme above
is also applicable when the proposal for $x'$ changes some or all of
the components of $x$. However, producing a multivariate candidate
$x'$ that has an appreciable chance of being accepted (i.e., satisfying
the inequality in line~\ref{alg:ssm:ratio} of Algorithm~\ref{alg:ssm})
while allowing appreciable movement, is
very difficult.  This highlights a very appealing aspect of the
single-site Metropolis scheme: even fairly thoughtless 1-d proposals
have an appreciable chance of being accepted while adequately
exploring the posterior.

There are clustering MCMC algorithms from statistical physics that
allow for many pixels in $x$ to be updated at once
\citep{Edwards&Sok:88}.  Such methods can be adapted to this
particular problem as in \citet{higd:1998a}, however such methods
typically show decreased efficiency relative to single-site
updating when the likelihood is strong relative to the prior.  This is
certainly the case with our attempts on this application whose 
results are not worth discussing here.
%
%
In the first edition of this handbook  \citet{higdon2011posterior} we also explored 
the {\em differential evolution}-MCMC (DE-MCMC)
sampler of \cite{braak2006mcm}.  However, we could not find an implementation that effectively sampled this EIT posterior so we do not further discuss DE-MCMC.
Here we'll consider some implementations of multivariate random walk Metropolis updating as competitors to the costly single-site Metropolis updating for our EIT application.

Later, in Section~\ref{sec:approx} we will also explore multivariate updates generated semi-automatically from the single-site Metropolis scheme by the MSDA algorithm, and highlight the \emph{adaptive} MSDA algorithm that tunes a modified likelihood function to improve efficiency by leveraging an otherwise inadequate approximation to the forward map. 

\subsection*{Random walk Metropolis} \label{sec:rwm}

The multivariate random walk Metropolis (RWM) has been the focus of a
number of theoretical investigations \citep{Tierney:94,gelman1996emj}.
But to date this scheme has not been widely used in applications, and
has proven advantageous only in simple, unimodal settings.  The
preference for single-site, or limited multivariate updates in
practice may be attributed to how the full conditionals often simplify
computation, or may be due to the difficulty in tuning highly
multivariate proposals. 
In our EIT application, the univariate full
conditionals do not lead to any computational advantages.
If there is ever an application for which RWM may be preferable, this
is it.  Single-site updating is very costly, and may be inefficient
relative to multivariate updating schemes for this multimodal
posterior.
The adaptive Metropolis (AM) method of~\cite{haario2006dram}
automatically tunes a multivariate Gaussian proposal, and allows 
relatively automatic sampling for EIT with GMRF prior~\citep{TAGHIZADEH2020112959}. 
However, we have not found that direct use of AM successfully samples in a complex inverse problem with
multimodal posterior~\citep[see, e.g.][]{cui2011bayesian}, 
such as the present posterior~\eqref{eq:graypost} that allows reconstruction of sharp conductivity boundaries in EIT.

A multivariate Gaussian random walk Metropolis scheme for
the $m$-vector $x$ is summarized in Algorithm \ref{alg:grwm}
using pseudo code.\\

\begin{algorithm}[h!]
\caption{Random Walk Metropolis}
\label{alg:grwm}
\begin{algorithmic}[1]
\Statex
\State initialize $x$
\For{$k=1:\text{niter}$}
  \State $x' = x + z$, where $z \sim N_m(0,\Sigma_z)$
  \If{ $u < \frac{\pi(x'|y)}{\pi(x|y)}$, where $u \sim U(0,1)$ }
     \State set $x = x'$
  \EndIf
\EndFor
\end{algorithmic}
\end{algorithm}

\noindent
We consider three different proposals for this scheme:
\begin{eqnarray*}
\Sigma_z &\propto& \Sigma_1 = I_m\\
\Sigma_z &\propto& \Sigma_2 = \mbox{diag}(s_1^2,\ldots,s_m^2)\\
\Sigma_z &\propto& \Sigma_3 = S^2
\end{eqnarray*}
where $s_i^2$ is the posterior marginal sample variance for the
conductivity $x_i$, and $S^2$ is the $m \times m$ sample covariance
matrix -- both estimated from the previously obtained single-site MCMC
run.  In each case we set $\Sigma_z = \alpha_i \Sigma_i$ where the
scalar $\alpha_i$ is chosen so that the candidate $x'$ is accepted
30\% of the time, which is close to optimal in a Gaussian setting.
\begin{figure}[h!t]
\centerline{
\begin{picture}(385,225)(0,0)
\put(20,20){\includegraphics[width=2.5in,angle=0]{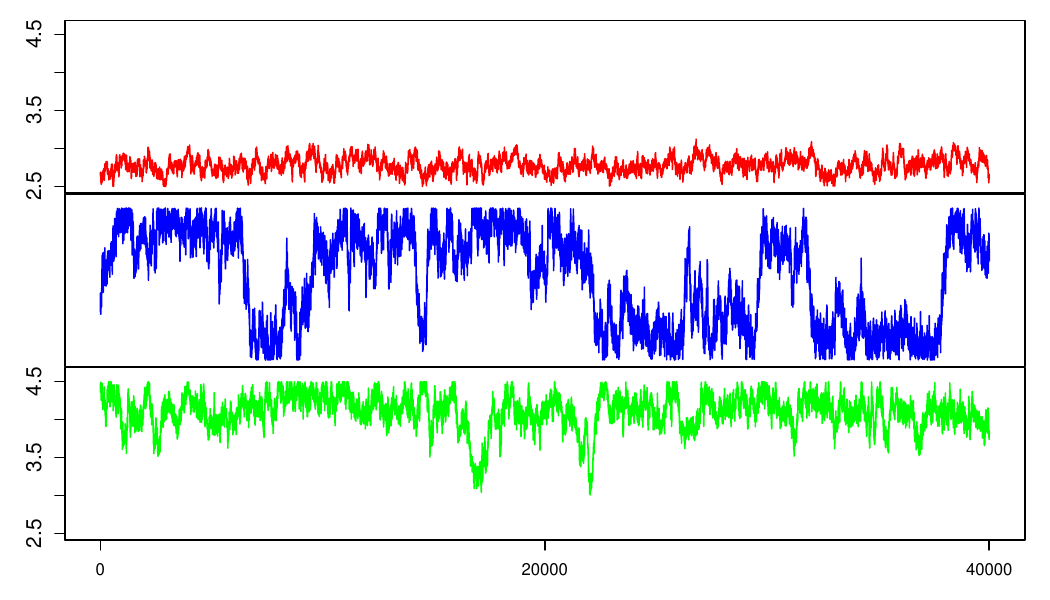}}
\put(195,20){\includegraphics[width=2.5in,angle=0]{ssm}}
\put(20,120){\includegraphics[width=2.5in,angle=0]{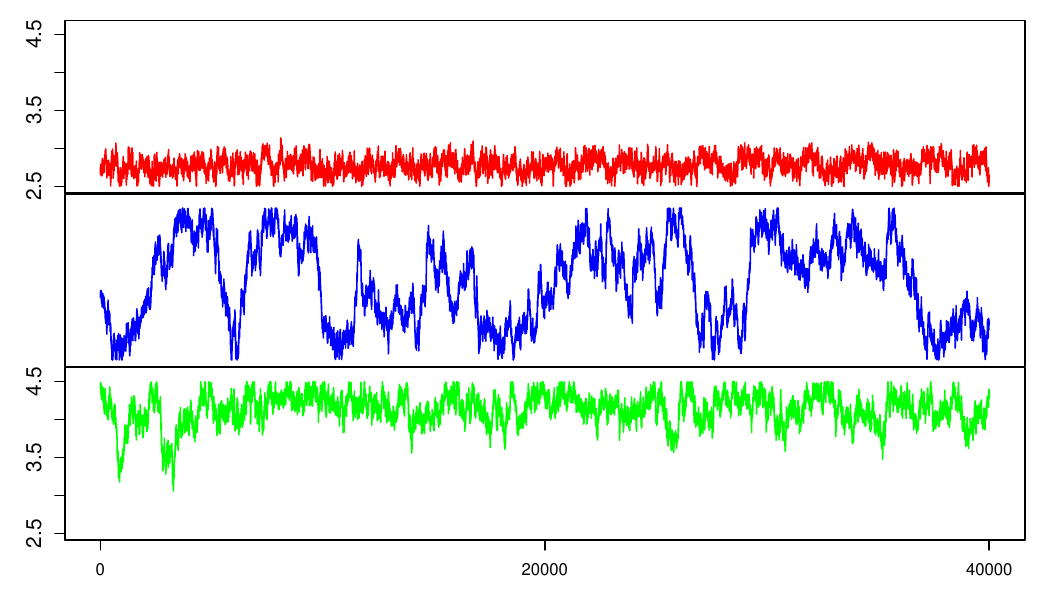}}
\put(195,120){\includegraphics[width=2.5in,angle=0]{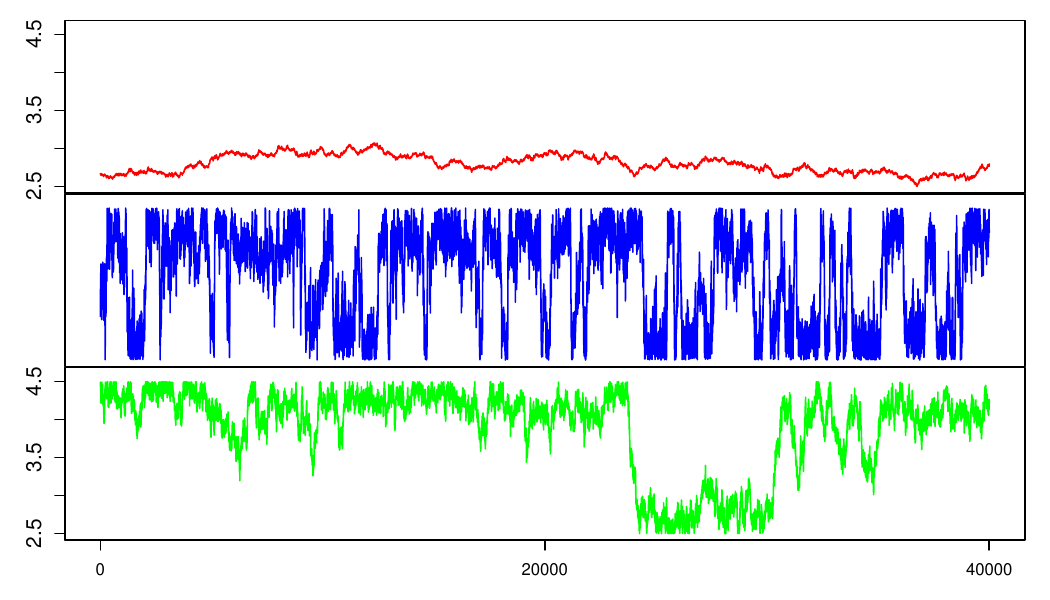}}
\put(120,210){\makebox(0,0){\sf \footnotesize RWM  $\Sigma_z \propto I$}}
\put(120,110){\makebox(0,0){\sf \footnotesize RWM  $\Sigma_z \propto$ diag$(s^2_1,\ldots,s^2_m)$}}
\put(290,210){\makebox(0,0){\sf \footnotesize RWM  $\Sigma_z \propto S^2$}}
\put(290,110){\makebox(0,0){\sf \footnotesize single-site Metropolis}}
\put(200,12){\makebox(0,0){\sf computational effort (simulator evaluations $\times m$)}}
\put(5,100){\rotatebox{90}{\sf conductivity}}
\end{picture} 
}
\caption{\label{fig:rwm} MCMC traces of three pixels circled in Figure
  \ref{fig:ssm} under three multivariate random walk Metropolis (RWM)
  schemes, and single-site Metropolis.  For each run, $40000 \times m$
  forward simulator evaluations are carried out.  While the RWM scheme
  with $\Sigma_z \propto S^2$ results in good movement for the central
  pixel, the movement of the top and bottom pixels are clearly
  inferior to that of single-site Metropolis.}
\end{figure}

MCMC traces for these three implementations of RWM are shown in Figure
\ref{fig:rwm}. The traces from the single-site Metropolis scheme are
also given for comparison.  Interestingly, the behavior of the traces
varies with the choice of $\Sigma_z$.  The scheme with $\Sigma_z
\propto S^2$ shows the most movement for the central pixel, which
moves between high and low conductivity over the run.  However its
performance for the top, low conductivity pixel is noticeably worse.
None of the RWM schemes do as well as single-site Metropolis when
looking at the bottom, high conductivity pixel.  These results suggest
that a scheme that utilizes both single-site and RWM updates with
$\Sigma_z \propto S^2$ might give slightly better posterior
exploration than single-site Metropolis alone.

\section{Augmenting with fast, approximate simulators} 
\label{sec:approx}

In many applications, a faster, approximate simulator is available
for carrying out the forward model evaluation $\eta_{\rm approx}(x)$.  
There are a number of approaches for using fast,
approximate simulators to improve the MCMC exploration of the original (fine-scale) posterior given in (\ref{eq:graypost}).  In this chapter we consider two related approaches: Metropolis-coupled MCMC  
\citep{geye:1991b,higd:lee:holl:2003,andersen2003big}; and delayed acceptance (DA) schemes that limit the
number of calls to the expensive, ``exact'' simulator 
(\citealp{liu2001mcs},\citealp{christen2005mcm},\citealp{cui2011bayesian},\citealp{banterle2019accelerating}).  
We'll also highlight the adaptive, multi-level extension of DA given in \citet{lykkegaard2023multilevel}, demonstrating it for the EIT application of this chapter.

We consider two different approximate forward models:
1) $\eta_\text{m}(x)$, based on an incomplete multigrid solve; and
2) $\eta_\text{c}(x)$, based on a coarsened representation of the conductivity
image $x$.
The solver used for $\eta_\text{m}(x)$ is derived from the BoxMG algorithm of \cite{JEDendy_1982a} and is described in \cite{higdon2011posterior}.  Evaluating $\eta_\text{m}(x)$ takes about a third of the time of evaluating the standard forward model $\eta(x)$.  It calculates rather accurate voltages; all well within $\sigma$ for each component of $\eta(x)$.  
In contrast, the coarse forward model $\eta_\text{c}(x)$ is well over 100 times faster that $\eta(x)$, but far less accurate.  
\begin{figure}[t!h]
   \centerline{
      \includegraphics[width=0.99\linewidth]{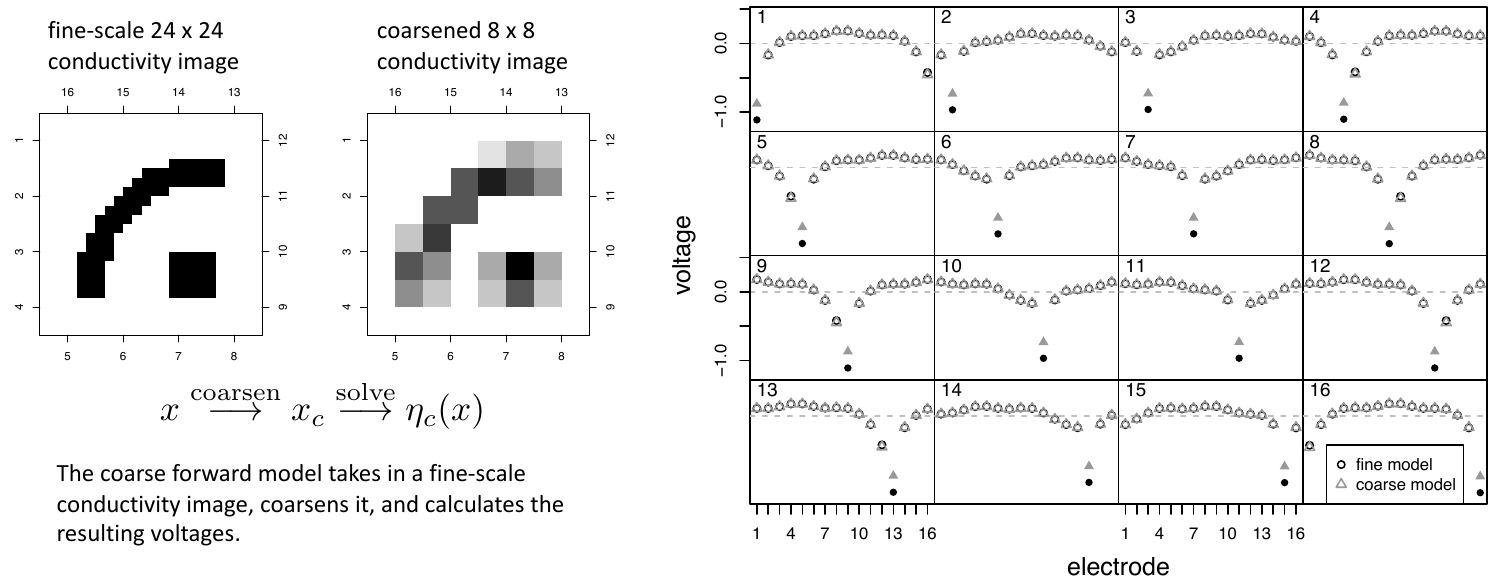}
   }
   \vspace*{-9pt}

   \caption{Fast, approximate coarse forward model $\eta_\text{c}(x)$.  The coarse forward
   model is produced by coarsening the $24 \times 24$ conductivity image (left) and then solving for the measured voltages (right).  For the fine-scale image $x$ (top left), the voltages produced by the
   original (fine-scale) forward model $\eta(x)$ are given by the circles; the voltages produced by the coarse-scale forward model $\eta_\text{c}(x)$ are given
   by the triangles.  The computed voltages differ substantially at the injector electrodes denoted by the filled in symbols.}
   \label{fig:coarseFine}
\end{figure}
Its calculated voltages can be well over $100\sigma$ away from $\eta(x)$ at the injector electrodes (see Figure \ref{fig:coarseFine}).  This error in $\eta_\text{c}(x)$ poses problems for the basic implementations of Metropolis coupling and delayed acceptance described below, motivating an adaptive error implementation described in Section \ref{sec:mlada}.

\subsection{Metropolis coupling}

\begin{figure}[t!h]
\centerline{
  \includegraphics[width=0.9\linewidth]{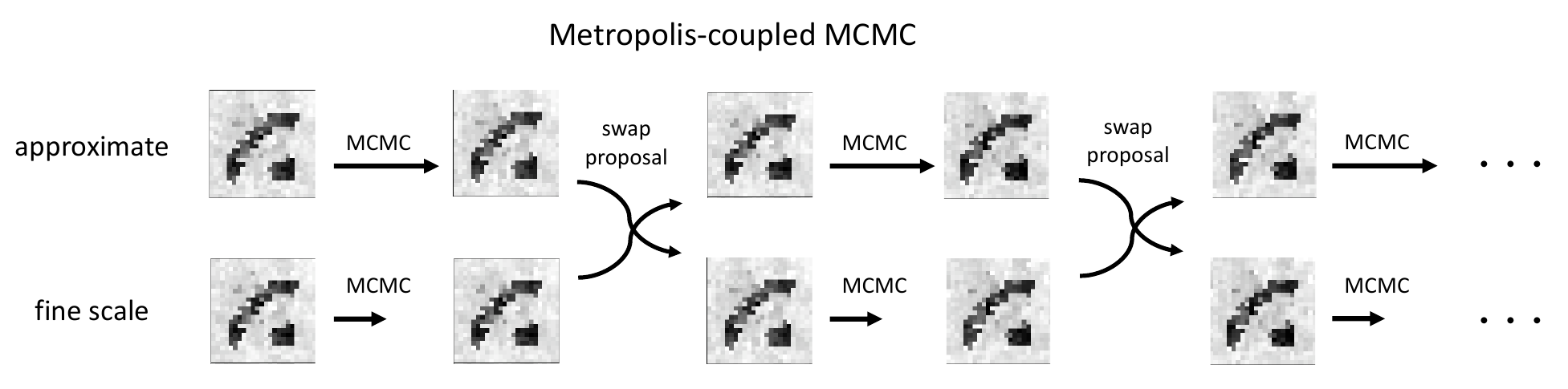}
}
\caption{A sequence from a Metropolis coupled sampler over a joint fine-scale and approximate posterior.   A scan of $3m$ single-site Metropolis 
updates are carried out on the approximate posterior while a scan of $m$ updates are carried out on the fine scale posterior.  This is followed by a Metropolis proposal that swaps the current images of the approximate and fine scale chains.  Marginally, the draws on the fine scale are from the original posterior distribution.
\label{fig:metCoupled} }
\end{figure}

By augmenting the posterior of interest with auxiliary distributions
one can use Metropolis coupling \citep{geye:1991b}, simulated
tempering \citep{mari:pari:1992}, or related schemes \citep{liu1999ssm}.
Here we use Metropolis coupling (MC) and define a joint, product
posterior for the pair of $24 \times 24$ conductivity images $(x,\tilde{x})$ given by
\begin{eqnarray*}
\pi(x,\tilde{x}|y) & = \pi(x|y) \times \pi_\text{m}(\tilde{x}|y) \propto & L(y|x) \pi(x) \times 
                    L_\text{m}(y|\tilde{x}) \pi(\tilde{x}), 
\end{eqnarray*}
where the same priors are specified for $x$ and $\tilde{x}$, and $L_\text{m}(y|\cdot)$ is the original likelihood, but with
the multigrid forward model $\eta_\text{m}(\cdot)$
\[
L_\text{m}(y|\tilde{x}) \propto 
\exp\left\{ -\frac{1}{2 \sigma^2} (y - \eta_\text{m}(\tilde{x}))^T  (y-\eta_\text{m}(\tilde{x})) \right\}.
\]
The posterior sampling proceeds by carrying out a fixed number of single site updates for $x$ and $\tilde{x}$, using their respective posteriors, and then proposing a swap between $x$ and $\tilde{x}$ as shown in Figure \ref{fig:metCoupled}.  The swap from $(x,\tilde{x})$ to $(\tilde{x},x)$ is a deterministic Metropolis proposal, accepted with probability 
$\mbox{min}\{1,\pi(\tilde{x},x|y) / \pi(x,\tilde{x}|y)  \}$. Since $\eta_\text{m}(x)$ can be evaluated in a third of the time required for $\eta(x)$, a parallel MC-MCMC implementation should allow three times the number of single-site Metropolis updates on the approximate posterior relative to the fine scale posterior before proposing a swap.  This will approximately balance the load between the two processors carrying out the single site updates.
Using the multigrid approximate forward model results in swaps that are accepted about 80\% of the time, owing to its fidelity to the fine scale forward model. If the coarse forward model $\eta_\text{c}(\cdot)$ is used, swaps are never accepted due to its large error.



\subsection{Delayed acceptance Metropolis}

The delayed acceptance approach of  \citet{christen2005mcm}
uses a fast, approximate simulator to ``filter'' a proposal, similarly to the surrogate transition method of \citet{liu2001mcs}, for dealing with complex forward
models.  For now, we define approximate posterior formulation using the multigrid simulator $\eta_\text{m}(x)$:
\begin{eqnarray*}
\pi_\text{m}(x|y) & \propto & L_\text{m}(y|x) \times \pi(x).
\end{eqnarray*}

A simple Metropolis-based formulation of this scheme is given 
in Algorithm \ref{alg:dam_mcmc},
\begin{algorithm}[hb!]
\caption{Delayed Acceptance Metropolis}
\label{alg:dam_mcmc}
\begin{algorithmic}[1]
\Statex
\State initialize $x$
\For{$k=1:\text{niter}$}
  \For{$i=1:m$}
    \State $x_i' = x_i + z$, where $z \sim N(0,\sigma^2_z)$ 
    \If{ $u_1 < \frac{\pi_\text{m}(x'|y)}{\pi_\text{m}(x|y)}$, where $u_1 \sim U(0,1)$  }
       \If {$u_2 < \frac{\pi(x'|y)\pi_\text{m}(x|y)}{\pi(x|y)
                   \pi_\text{m}(x'|y)}$, where $u_2 \sim U(0,1)$
               }
         \State set $x_i = x_i'$
       \EndIf
    \EndIf
  \EndFor
\EndFor
\end{algorithmic}
\end{algorithm}
where $\pi(x|y)$ and $\pi_\text{m}(x|y)$ denote the posteriors using the exact and
approximate simulators respectively.  Notice that the exact simulator need only
be run if the filtering condition ($u_1 < \frac{\pi_\text{m}(x'|y)}{\pi_\text{m}(x|y)}$) 
involving the faster, approximate simulator is
satisfied.  Hence, if the proposal width is chosen so that the filtering condition is
satisfied only a third of the time, the exact simulator is only run for a third of
the MCMC iterations.  Using the multigrid simulator $\eta_\text{m}(x)$, 
the $40000 \times m$ iterations required for our original single-site Metropolis
scheme take about 66\% of the computational effort using this delayed acceptance approach.  
Using the coarse simulator $\eta_\text{c}(x)$ in this delayed acceptance algorithm is ineffective.  Any proposal accepted in the approximate test (step 5, algorithm \ref{alg:dam_mcmc}) is nearly always rejected in the fine scale test (step 6).
\begin{figure}[t!h]
\centerline{
  \includegraphics[width=0.9\linewidth]{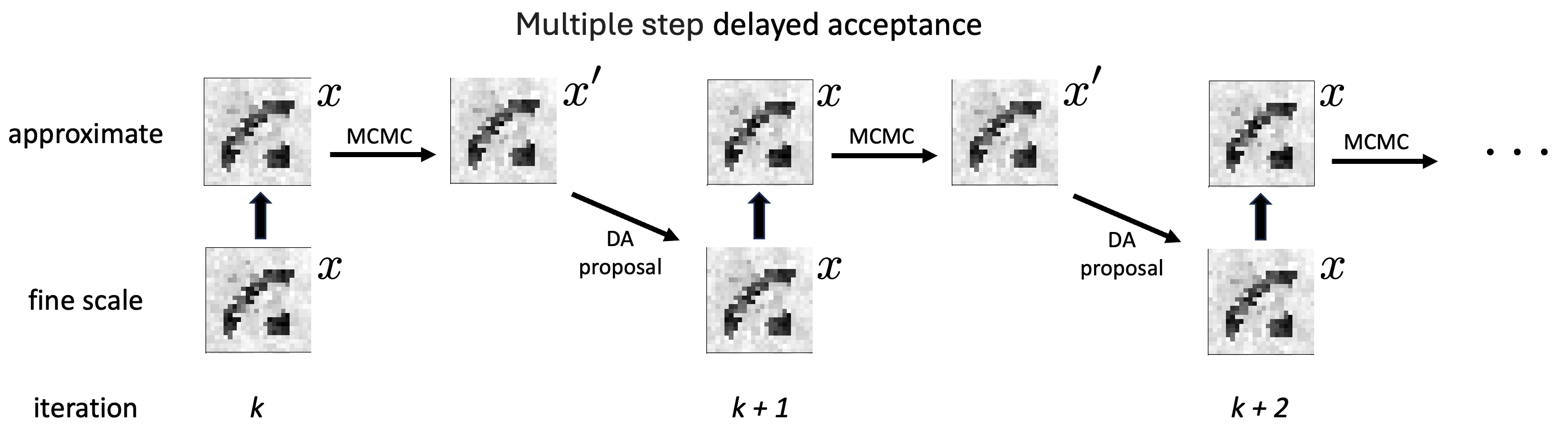}
}
\caption{Two($+$) updates from a multiple step delayed acceptance (MSDA) sampler. The current state $x$ from the fine scale is moved up ($\boldsymbol{\uparrow}$) to the approximate level.  This serves as the starting point for one (or more) single site Metropolis updates carried out according to the approximate posterior $\pi_\text{m}(x|y)$ depicted by $\stackrel{\mbox{\tiny MCMC}}{\longrightarrow}$.  The resulting candidate $x'$ is accepted on the fine scale ($\searrow$) with probability $\mbox{min}\{1,\pi(x'|y)\pi_\text{m}(x|y)/\pi(x|y)\pi_\text{m}(x'|y)$.  If not accepted, the chain remains at $x$ for this DA step.
\label{fig:mlda} }
\end{figure}

\noindent
{\bf Multiple step delayed acceptance} Algorithm \ref{alg:dam_mcmc} ``filters'' each proposal (step 4) before computing an acceptance outcome (step 6).  Alternatively, multiple MCMC steps may be taken on the approximate formulation to produce a candidate $x'$ that differs from the current state $x$ at more than just a single site $i$.
\begin{algorithm}[ht!]
\caption{Multiple Step Delayed Acceptance Metropolis}
\label{alg:msdam_mcmc}
\begin{algorithmic}[1]
\Statex
\State initialize $x$
\For{$k=1:\text{niter}$}
  \State set $x' = x$
  \For{$j=1:n_{\rm step}$}
    \State $x_i^* = x_i + z$, where $z \sim N(0,\sigma^2_z)$ and $i \sim U\{1,\ldots,m\}$ 
    \If{ $u_1 < \frac{\pi_\text{m}(x^*|y)}{\pi_\text{m}(x'|y)}$, where $u_1 \sim 
       U(0,1)$  } 
       \State $x_i' = x^*_i$
    \EndIf
  \EndFor
  \If {$u_2 < \frac{\pi(x'|y)\pi_\text{m}(x|y)}{\pi(x|y)
                   \pi_\text{m}(x'|y)}$, where $u_2 \sim U(0,1)$
               }
         \State set $x = x'$
  \EndIf
\EndFor
\end{algorithmic}
\end{algorithm}
This multiple step delayed acceptance (MSDA) is detailed in Algorithm~\ref{alg:msdam_mcmc}.  This algorithm produces updates that are in detailed balance with the fine scale posterior $\pi(x|y)$ \citep{liu2001mcs,lykkegaard2023multilevel}.  Figure \ref{fig:mlda} depicts this algorithm; now the horizontal arrows ($\stackrel{\mbox{\tiny MCMC}}{\longrightarrow}$) denote $n_{\rm step}$ single site MCMC steps using the approximate posterior $\pi_\text{m}(x|y)$ as the target distribution.
%
%
The collection of multiple steps needs to define a Markov kernel that is in detailed balance with the  approximate posterior $\pi_\text{m}(x|y)$ for the second acceptance probability (step 10 in Alg.~\ref{alg:msdam_mcmc}) to be correct and give an algorithm that produces draws from $\pi(x|y)$.  Randomly choosing the site to update (step 5, Alg.~ \ref{alg:msdam_mcmc}) achieves this, whereas the deterministic scan in Algorithm~\ref{alg:ssm} does not.

This MSDA algorithm is very similar to one half of Metropolis coupling, by following a single flow of arrows in Figure~\ref{fig:metCoupled}  -- the acceptance of the swap proposal in MC-MCMC is identical to acceptance of the eventual proposal $x'$ produced by MSDA.  However, there are major differences.  MC-MCMC requires stationarity on the joint product distribution $\pi(x|y) \times \pi_\text{m}(\tilde{x}|y)$.  MSDA aims for stationarity only on the fine scale posterior $\pi(x|y)$.  The moves made according to the approximate posterior $\pi_\text{m}(x|y)$ effectively produce a multivariate proposal $x'$ to be considered at the fine scale.  If $n_{\rm step}$ is small, the proposal $x'$ is ``close'' to the fine scale posterior; if $n_{\rm step}$ very large, the proposal $x'$ is closer to the approximate posterior.  Hence, $n_{\rm step}$ could serve as an MCMC tuning parameter, much like the proposal width used for the single site Metropolis steps.  For the example in Section \ref{sec:mlada}, we take $n_{\rm step} = 100$.

Finally, we note that \citet{christen2005mcm} and \citet{fox2020randomized} give a more general formulation for the delayed acceptance sampler for which the approximate simulator can depend on the current state $x$ of the chain; the computed example in \citet{christen2005mcm} used a state-dependent linearization of the forward map.  While a bit more demanding computationally, the more general algorithm can gain further efficiency by making use of local approximations and error models.
Additionally, we remark that while we used a coarsened grid to construct the approximate simulator, \textit{any} computationally cheaper approximation can be employed. Depending on the broader context of the problem, options include reduced-precision computing \citep{higham_mixed_2022}, early stopping of iterative solvers \citep{wikle_spatiotemporal_2001}, early termination of time-dependent problems \citep{lykkegaard2023multilevel}, dataset subsampling \citep{quiroz_speeding_2019}, reduced grid resolution \citep{lykkegaard2023multilevel}, surrogate models \citep{laloy_efficient_2013,seelinger_democratizing_2024}, and partially-computed likelihoods~\citep{banterle2019accelerating}.

\subsection{Adaptive, multiple step delayed acceptance}
\label{sec:mlada}
\begin{figure}[h]
   \centerline{
      \includegraphics[width=0.7\linewidth]{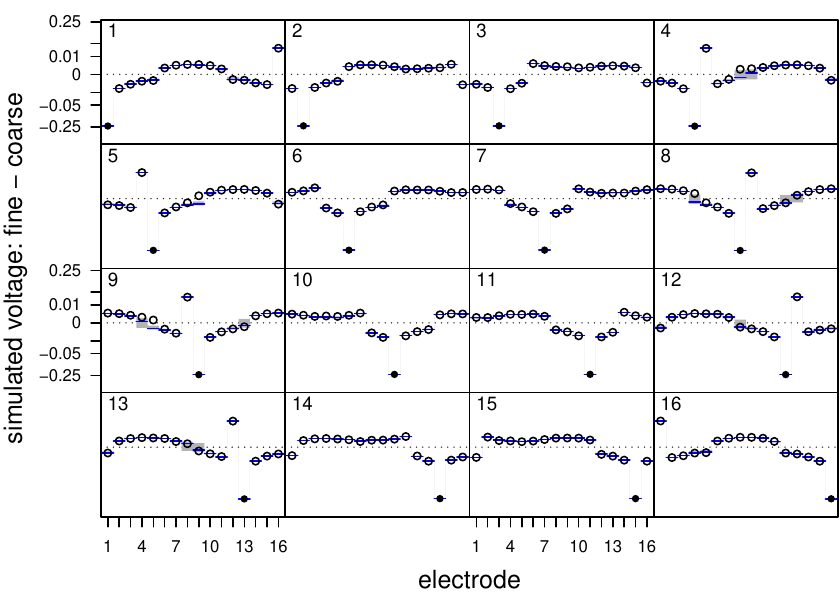}
   }
   \vspace*{-9pt}

   \caption{The difference between the fine scale simulator $\eta(x)$ and the coarse simulator $\eta_\text{c}(x)$.  Here $\eta(x)-\eta_\text{c}(x)$ is shown ($\circ$ plotting symbols) for the image $x$ given in the left hand frame of Figure \ref{fig:eit}.  The differences are consistently large and negative at the injector electrodes marked by the solid circles.  The adaptively estimated bias term is also shown (- plotting symbols). }
   \label{fig:diffData}
\end{figure}

So far, the coarse $8\times 8$ solver-based forward model has been ineffective in these MC-MCMC or MSDA implementations, in spite of its over 100-fold speed up in evaluating $\eta_\text{c}(x)$.  This is because the coarse scale simulator $\eta_\text{c}(x)$ differs too much from the fine scale simulator $\eta(x)$; simply swapping out the fine with the coarse simulator in $L_\text{c}(y|x)$ results in a posterior that is too far away from the target $\pi(x|y)$.  Figure \ref{fig:diffData} shows the difference $\eta(x) - \eta_\text{c}(x)$ for the $24 \times 24$ conductivity field shown in the left hand frame of Figure \ref{fig:coarseFine}.  The difference is largest at the injector electrodes.

\begin{figure}[t!h]
\centerline{
\begin{picture}(385,120)(0,0)
\put(18,20){\includegraphics[width=2.5in,angle=0]{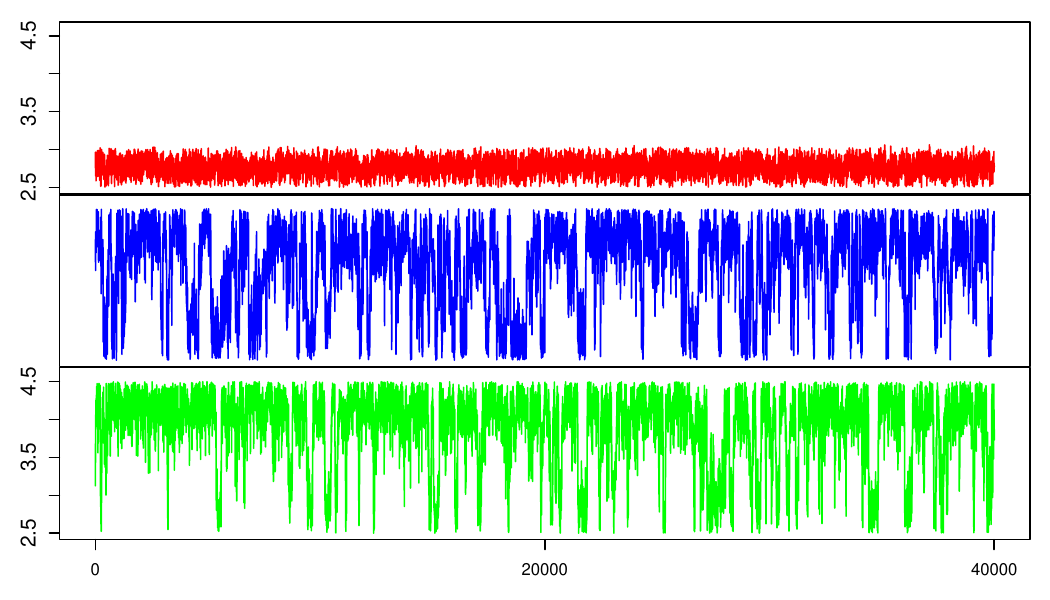}}
\put(195,20){\includegraphics[width=2.5in,angle=0]{ssm}}
%
\put(112,110){\makebox(0,0){\sf \footnotesize adaptive, multiple step delayed acceptance}}
\put(290,110){\makebox(0,0){\sf \footnotesize single-site Metropolis}}
%
%
\put(200,12){\makebox(0,0){\sf computational effort (simulator evaluations $\times m$)}}
\put(5,45){\rotatebox{90}{\sf conductivity}}
\end{picture} 
} 
\caption{\label{fig:amsda} MCMC traces of three pixels circled in Figure
  \ref{fig:ssm} using the adaptive, multiple step delayed acceptance (AMSDA) Metropolis scheme.  For comparison, the traces under the single site Metropolis scheme are shown.  The $x$-axis marks computational effort, which is the same for the two MCMC efforts.  The AMSDA sampler shows 
  substantial improvement, making effective use of the very fast, but highly biased, coarse scale model.}
\end{figure}

Accounting for bias and error due to using an approximate model in
place of the accurate, fine scale model has been a focus in Bayesian inverse problems \citep{kaipio2007statistical} and in computer model emulation \citep{kenn:ohag:2000}.  Here we follow \citet{cui2011bayesian} and \citet{fox2020randomized} and use adaptive MCMC 
\citep{roberts2009examples,roberts2007coupling} to adapt a ``coarse" likelihood $L_\text{c}(y|x)$ so that multiple step proposals $x'$ are more readily accepted in DA.  At step $k$, $L_\text{c}(y|x)$ is modified with an adaptive bias $16^2$-vector $b_k$ and a $16^2 \times 16^2$ covariance matrix $\Sigma_{bk}$ 
\[
L_{\text{c}k}(y|x) \propto \exp\left\{ -\half (y - \eta_\text{c}(x) - b_k)^T
(\sigma^2 I + \Sigma_{bk})^{-1} (y - \eta_\text{c}(x) - b_k)
\right\}.
\]
This gives an adaptive version of the coarse scale posterior  $\pi_{\text{c}k}(x|y) = L_{\text{c}k}(y|x) \pi(x)$.  This adaptive, multiple step delayed acceptance algorithm is described in Algorithm 
\ref{alg:amsda}.
%
\begin{algorithm}[th!]
\caption{Adaptive, Multiple Step Delayed Acceptance Metropolis}
\label{alg:amsda}
\begin{algorithmic}[1]
\Statex
\State initialize $x$, $b_0=\eta(x)-\eta_\text{c}(x)$ and $\Sigma_{b0} = \mathbf{0}$
\For{$k=1:\text{niter}$}
  \State set $x' = x$
  \For{$j=1:n_{\rm step}$}
    \State $x_i^* = x_i + z$, where $z \sim N(0,\sigma^2_z)$ and $i \sim U\{1,\ldots,m\}$ 
    \If{ $u_1 < \frac{\pi_{\text{c}k-1}(x^*|y)}{\pi_{\text{c}k-1}(x'|y)}$, where $u_1 \sim 
       U(0,1)$  } 
       \State $x_i' = x^*_i$
    \EndIf
  \EndFor
  \If {$u_2 < \frac{\pi(x'|y)\pi_{\text{c}k-1}(x|y)}{\pi(x|y)
                   \pi_{\text{c}k-1}(x'|y)}$, where $u_2 \sim U(0,1)$
               }
         \State set $x = x'$
  \EndIf
  \State $b_k = \frac{1}{k}\left[ (k-1)b_{k-1} + \eta(x)-\eta_\text{c}(x) \right]$
  \State $\Sigma_{bk} = \frac{1}{k}\left[ (k-1)\Sigma_{bk-1} + (\eta(x)-\eta_\text{c}(x)-b_k)(\eta(x)-\eta_\text{c}(x)-b_k)^T \right]$
\EndFor
\end{algorithmic}
\end{algorithm}
Our implementation takes $n_{\rm step} = 100$, $\sigma_z = 0.3$, and initializes $b_0=\eta(x)-\eta_\text{c}(x)$ and $\Sigma_{b0}=\bf{0}$.
%
Figure \ref{fig:amsda} shows the MCMC histories for the same three conductivity pixels shown in previous figures.  The computational effort is equivalent to that of Figure \ref{fig:ssm} -- the equivalent of 40,$000 \times m$ forward model runs on the fine scale.  After many iterations, the adaptive components $b_k$ and $\Sigma_k$ stabilize; the resulting values for $b_\infty$ are given by the dash (-) symbols in Figure \ref{fig:diffData}.  

\noindent
We employed the free and open-source Python software library \texttt{tinyDA} \citep{tinyDA} for the numerical experiments involving Delayed Acceptance MCMC outlined here. In addition to standard Metropolis-Hastings, DA and MSDA MCMC, \texttt{tinyDA} provides routines for Multilevel Delayed Acceptance (MLDA), a multilevel extension to MSDA, which enables MCMC sampling using a multilevel hierarchy of increasingly coarser models \citep{lykkegaard2023multilevel}. When carefully implemented, such a multilevel model hierarchy can provide even higher gains to the MCMC sampling efficiency than the two-level sampler employed in this work.

%

\section{Discussion}

For the EIT example, single-site Metropolis requires about 20 million
simulator evaluations to effectively sample this posterior
distribution.  Multivariate updating schemes such as random walk
Metropolis or DE-MCMC -- as we implemented them here, or in the previous version of this chapter -- don't offer
any real relief.  Tempering and Metropolis-coupling schemes can help if the approximate forward model is very close to the fine scale forward model.  Here the multigrid-based approximation is quite accurate, but is only three times faster.  Hence the speed-up with such a model will be of a similar magnitude. The AMSDA approach using the much faster (over $100\times$) coarse forward model seems to be an exception.  The combination of the adaptive error term and the multiple steps before testing the candidate with the expensive, fine-scale forward model results in an MCMC scheme that is over an order of magnitude more efficient.  The success of this approach with other challenging inverse problems 
\citep{fox2020randomized,lykkegaard2023multilevel}
suggests this isn't just due to the special features of this particular EIT example.  We also note that for ease of presentation, we have described rather basic implementations of delayed acceptance schemes.  We point the reader to \citet{fox2020randomized}, \citet{banterle2019accelerating} and \citet{lykkegaard2023multilevel} for richer implementations.

One challenging feature of this application is the multimodal nature
of the posterior which is largely induced by our choice of prior.  By
specifying a more regularizing prior, such as the GMRF \ref{eq:gmrf})
or the process convolution (\ref{eq:conv}), the resulting posterior
will more likely be unimodal, so that standard MCMC schemes will be more
efficient.  Of course, the sacrifice is that one is now less able to
recover small-scale structure that may be present in the inverse
problem.

We considered two different fast, approximate simulators in this chapter.  There is a rather vast literature developing fast approximations to computationally intensive models.  This includes reduced order models \citep{BennerCOW2017ModRedApproxTA}, cross approximations \citep{dolgov2019hybrid}, and response surface/regression approaches such as polynomial chaos \citep{xiu2002wiener}, Gaussian processes \citep{kenn:ohag:2001}, or additive regression trees \citep{pratola2016bayesian}.  The response surface approaches are most useful when the effective dimension of the parameter vector is fairly small ($<20$) and the output response to parameter changes is smooth and predictable.  The response surface is typically estimated from a collection of model runs carried out at different input parameter settings.  By contrast, a well designed reduced order model can enable MCMC to handle some rather complicated high-dimensional, non-linear problems -- see \citet{keating2010optimization}, for example.  A comparison of reduced model and response surface approaches can be found in \citet{frangos2010surrogate}.

Finally we note that the traditional way to speed up the computation required to
solve an inverse problem is to speed up the simulator $\eta(x)$.  A substantial
amount of progress has been made to create simulators 
that run on highly distributed computing machines.
While MCMC does not lend itself to easy parallelization, we are seeing advances in the use of parallelized MCMC for inverse problems \citep{craiu2009learn,vrugt2016markov,nishihara2014parallel,conrad2018parallel,brockwell2006parallel}.
The integration of modern computing architecture with
MCMC methods will certainly extend the reach of MCMC based solutions to 
inverse problems.


\section*{Acknowledgments}
DH was supported, in part, by the U.S. Department of Energy, Office of Science, Office of Advanced Scientific Computing Research and Office of High Energy Physics, Scientific Discovery through Advanced Computing (SciDAC) program. CF is grateful to digiLab for support and a pleasant working environment.

\bibliographystyle{apalike}
\bibliography{./dave,./colin,./mikkel}

\begin{appendix}
\section{Formulation based on a process convolution prior}
An alternative to treating each pixel in the image as a parameter to
be estimated is to use a lower dimensional representation for the
prior.  Here we describe a process convolution \citep{higd:2002} prior
for the underlying image $x$.

\begin{figure}[t!h]
  \centerline{
   \includegraphics[width=4.0in,angle=0] {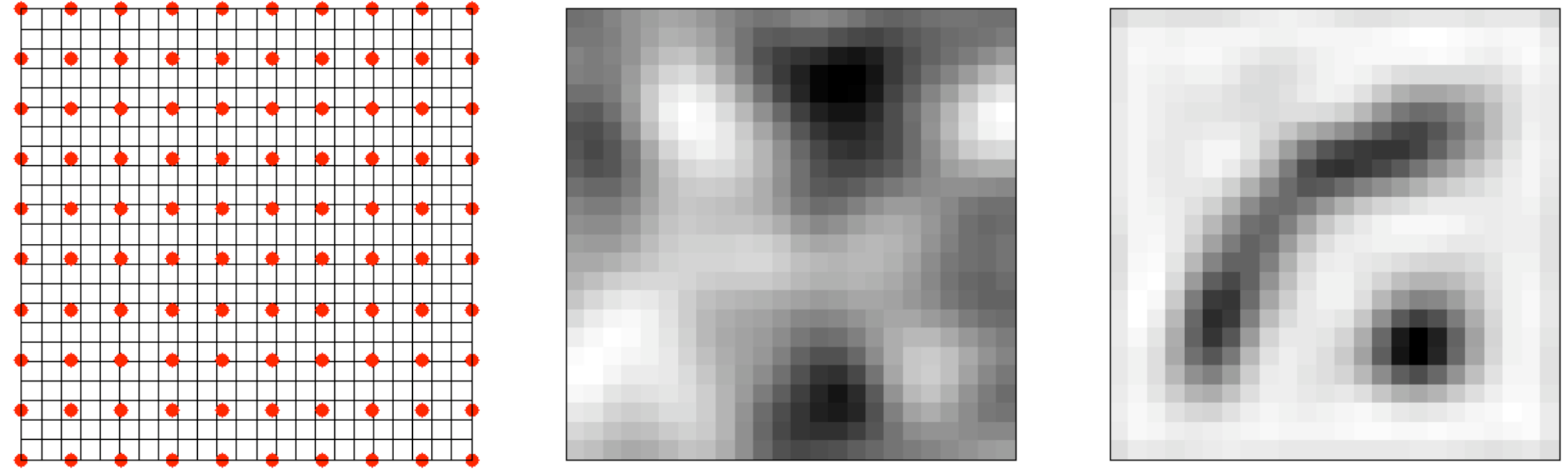}
  }
   \caption{\label{fig:bumps} Left: A $10 \times 10$ lattice of
   locations $\omega_1,\ldots,\omega_p$, for the $u_j$'s of
   the process convolution prior; the $24 \times 24$ image pixels are shown for reference.
   Middle: a realization from the process convolution prior for $x(s)$.
   Right: posterior mean from the single-site Metropolis scheme
   run on the $u$ vector controlling that controls the image $x$. }
\end{figure}

We define $x(s),\,s \in \Omega$ to be a mean zero Gaussian
process. But rather than specify $x(s)$ through its covariance
function, it is determined by a latent process $u$ and a
smoothing kernel $k(s)$.  The latent process $u = (u_1,\ldots,u_p)^T$ is
located at 
the spatial sites $\omega_1, \ldots, \omega_p$, also in
$\Omega$ (shown in Figure \ref{fig:bumps}).  The
$u_j$'s are then modeled as independent draws from a $N(0,\sigma^2_u)$
distribution.  The resulting continuous Gaussian process model for $x(s)$ is then
\begin{equation}
\label{eq:conv}
x(s) = \sum_{j=1}^p u_j k(s-\omega_j)
\end{equation}
where $k(\cdot - \omega_j)$ is a kernel centered
at $\omega_j$.
For the EIT application, we define the smoothing
kernel $k(\cdot)$ to be a
radially symmetric bivariate Gaussian density, with standard deviation
$\sigma_u = .11$.  Figure \ref{fig:bumps} shows a prior draw from this
model over the $24 \times 24$ pixel sites in $\Omega$.
Under this formulation, the image $x$ is controlled by $p=100$
parameters in $u$.  Thus a single-site Metropolis scan of $u$ takes
less than 20\% of the computational effort required to update each
pixel in $x$.  In addition, this prior enforces very smooth
realizations for $x$.  This makes the posterior distribution better
behaved, but may make posterior realizations of $x$ unreasonably
smooth.  The resulting posterior mean for $x$ is shown in Figure
\ref{fig:bumps}.  For a more detailed look at process convolution
models, see \citep{higd:2002};  \citet{Paci:Sche:2004} give non-stationary
extensions of these spatial models.
\end{appendix}

\end{document}